\documentclass[longauth]{aa}
\usepackage{rotating}
\usepackage{natbib}
\usepackage{lscape}
\usepackage{color}
\usepackage{graphicx}

\begin{document}

\title{Isotopic ratios in outbursting comet C/2015 ER61}

\author{
Bin Yang  \inst{1,2}\and 
Damien Hutsem{\'e}kers \inst{3}\and 
Yoshiharu Shinnaka \inst{4}\and 
Cyrielle Opitom \inst{1}\and 
Jean Manfroid \inst{3}\and 
Emmanu{\"e}l Jehin \inst{3}\and 
Karen J. Meech \inst{5}\and 
Olivier R. Hainaut \inst{1}\and 
Jacqueline V. Keane \inst{5}\and 
Micha{\"e}l Gillon\inst{3}
}

\institute{
  European Southern Observatory, Alonso de C\`{o}rdova 3107,  Vitacura, Casilla 19001, Santiago, Chile \and
  Yunnan Observatories, Chinese Academy of Sciences, Kunming 650011, China \and
  Institut d'Astrophysique et de G\'eophysique, Universit\'e de Li\`ege, 19c All\'ee du Six Ao\^ut, B-4000 Li\`ege, Belgium \and
  National Astronomical Observatory of Japan, 2-21-1 Osawa, Mitaka, Tokyo 181-8588, Japan
  \and
  Institute of Astronomy, University of Hawaii, 2680 Woodlawn Dr., Honolulu, HI 96822, USA
}

\abstract{Isotopic ratios in comets are critical to understanding the origin of cometary material and the physical and chemical conditions in the early solar nebula. Comet C/2015 ER61 (PANSTARRS) underwent an outburst with a total brightness increase of 2 magnitudes on the night of 2017 April 4. The sharp increase in brightness offered a rare opportunity to measure the isotopic ratios of the light elements in the coma of this comet. We obtained two high-resolution spectra of C/2015 ER61 with UVES/VLT on the nights of 2017 April 13 and 17. At the time of our observations, the comet was fading gradually following the outburst. We measured the nitrogen and carbon isotopic ratios from the CN violet (0,0) band and found that $^{12}$C/$^{13}$C=100 $\pm$ 15, $^{14}$N/$^{15}$N=130 $\pm$ 15. In addition, we determined the $^{14}$N/$^{15}$N ratio  from four pairs of NH$_2$ isotopolog lines and measured $^{14}$N/$^{15}$N=140 $\pm$ 28. The measured isotopic ratios of C/2015 ER61 do not deviate significantly from those of other comets.}


\maketitle

\section {Introduction}
Understanding how planetary systems form from protoplanetary disks remains  one of the great challenges in astronomy. In our own solar system,  a wealth of processes happened involving chemistry and dynamics
at all scales before reaching its present state. Comets are among the most unaltered materials in the solar system, having preserved their primitive element abundances. Comets witnessed the formation processes of the solar system and retain chemical signatures of the solar nebula in which they formed. Among all the observable properties, isotopic abundances are key tracers for reconstructing the origin and evolution of cometary material. Isotopic fractionation is sensitive to nebular environmental conditions, such as temperature, density, radiation, and composition \citep{Bockelee-Morvan:2015}. However, because of the faint signatures of isotopologs, isotopic measurements are only obtained in a handful of bright comets.

Discovered by the Pan-STARRS1 telescope on Haleakala on 2015 March 14, comet C/2015 ER61 (PANSTARRS, hereafter ER61) is on a highly eccentric orbit ($e=0.9973$, $a=385$ au), suggesting its origin is in the inner Oort cloud \citep{Meech:2017}. At the time of discovery, ER61 appeared asteroidal with no visible coma detected around the nucleus and it was thought to be a potential so-called Manx comet, which is an observationally inactive or nearly inactive object that is coming in from the Oort cloud \citep{Meech:2016}. The significance of Manx objects is that they may represent volatile-poor material that formed in the early inner solar system and was subsequently ejected to the Oort cloud during the solar system formation \citep{Meech:2016}. The comet became visually active a few months after its initial discovery with the appearance of a faint coma, which was detected by the Gemini telescope in 2015 June when the comet was 7.7 au from the sun. \citep{Meech:2017}. 
\begin{figure}[t]
\includegraphics[angle=90,width=9cm]{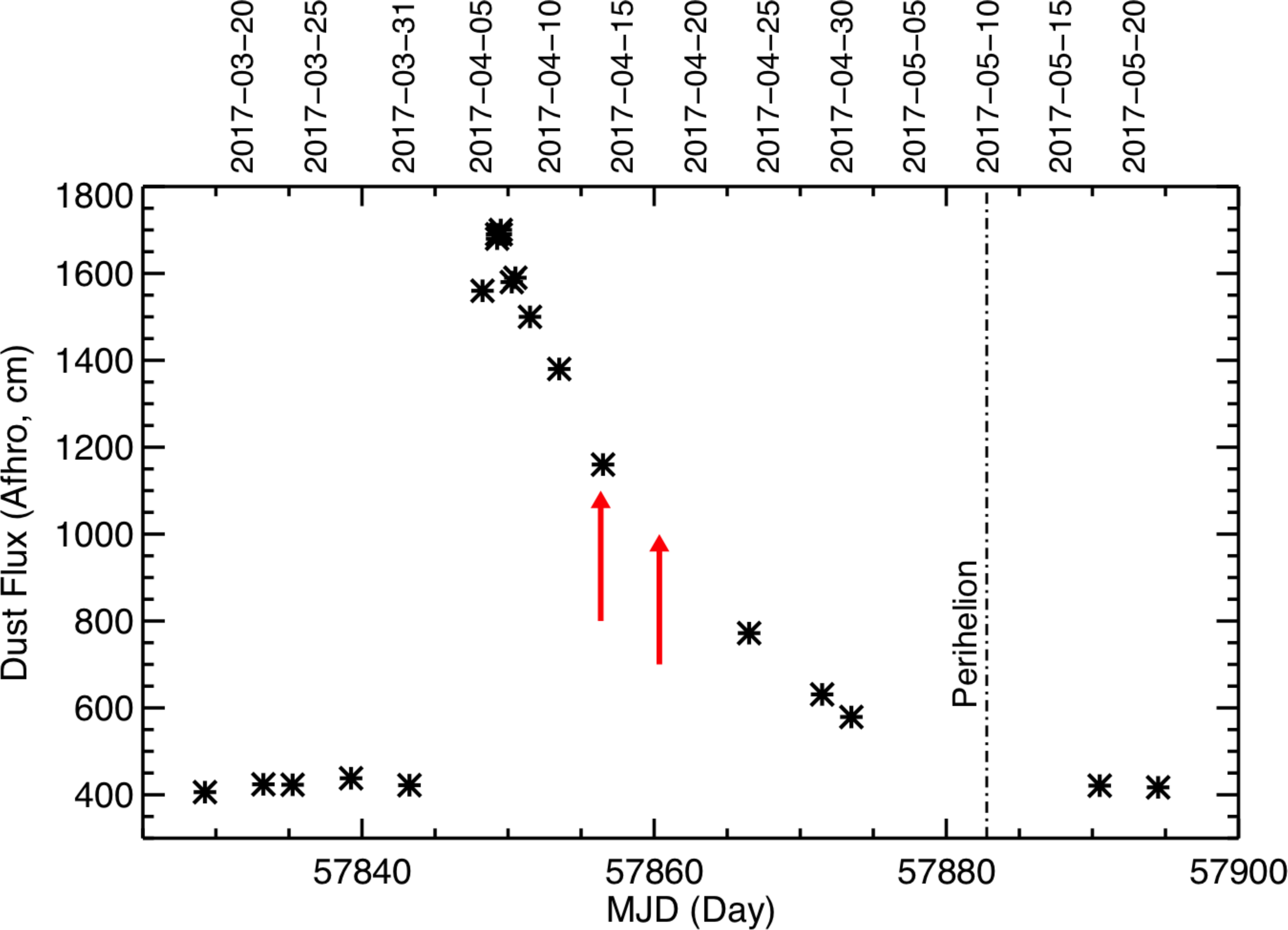}
\caption{Pre- and post-outburst evolution of the brightness of comet C/2015 ER61. The stars are the $Af\rho$ values we measured using I-band images obtained with the TRAPPIST-South telescope. The $Af\rho$ value is a proxy for the dust production rate \citep{ahearn:1984}. The two red arrows indicate the dates (UT 2017 April 13 and 17) of the UVES observations, which were just one week after the outburst. The vertical dashed line indicates the time when the comet reached its perihelion.}
\label{lcv}
\end{figure} 

On 2017 April 4, ER61 underwent a significant outburst, with a visual magnitude of 6.2, up from a pre-outburst brightness of 8.4 mag \footnote[1]{http://195.209.248.207/en/observation/listObserv/1383}. Photometric observations with the 60 cm TRAPPIST-South telescope \citep{Jehin:2011} at La Silla on 2017 April 5 show that the gas production rates increased by a factor of 7 compared to the previous observations made on 2017 March 31 \citep{Jehin:2017}, and the dust mass-loss rate increased at least by a factor of 4; see Figure 1. 

The outburst of ER61 provides a rare opportunity to study interior materials, excavated and released to the coma, which have been well protected from surface alterations (i.e., space weathering and cometary activity). Most importantly, the enhanced brightness of ER61 has enabled measurements of isotopic ratios of several species in the coma of this comet via high-resolution spectroscopy. High resolution is needed for isotopic studies because the emission lines of isotopes are weak and they have to be distinguished from emission lines of other isotopes (e.g., NH$_2$ lines are blended with the C$_2$ Swan lines) and from the underlying dust-scattered solar continuum \citep{Manfroid:2009}. 

More than a month after the outburst, ER61 reached its perihelion on 2017 May 9.94 at 1.042 au from the Sun. Later in 2017 June, a fragment co-moving with the nucleus was detected by the 16-inch Tenagra robotic telescope (MPEC 2017-M09). This newly detected secondary is now officially named C/2015 ER61-B (PANSTARRS). The outburst that occurred in early 2017 April may be associated with this splitting event, but the cause of the fragmentation has not been investigated and is beyond the scope of this paper. 

\section{Observation and data analysis}
Observations of comet ER61 were carried out in service mode with the Ultraviolet-Visual Echelle Spectrograph (UVES) mounted on the 8.2 m UT2 telescope of the European Southern Observatory. Using director's discretionary time, the total 10,800s of science exposure was divided into two exposures of 6400s each
on 2017 April 13 and 17, respectively. As shown in Figure 1, the outburst was short-lived. At the time of the UVES observations, the brightness of the comet was decreasing steadily, but was still much higher than the pre-outburst level. We used the atmospheric dispersion corrector and the UVES standard setting 
346 + 580 with dichroic $\#$1 that covers roughly from 3000 to 3880 \AA\ on the blue CCD and from 4760 to 6840 \AA\ on the two red CCDs. We used a 0.5 $\times$ $10\farcs 0$ slit, providing a resolving power R $\approx$ 80,000. 

The raw spectral data were reduced using the UVES Common Pipeline Library (CPL) data reduction pipeline \citep{Ballester:2000}, modified to accurately merge individual orders into a two-dimensional spectrum. Subsequently, the echelle package of the IRAF software was used to calibrate the spectra and to extract one-dimensional spectra. In turn, the cosmic rays were removed and the comet spectra were rebinned and corrected for the velocity of the comet. Lastly, the continuum component, including the sunlight reflected by cometary dust grains and the telluric absorption features, was removed. The final comet spectrum contains the gas component only, meaning the emissions from photodissociated radicals. More details regarding data reduction procedures are described in \cite{Manfroid:2009} and \cite{Shinnaka:2016}. 

\begin{figure}[!h]
\includegraphics[width=8.5cm]{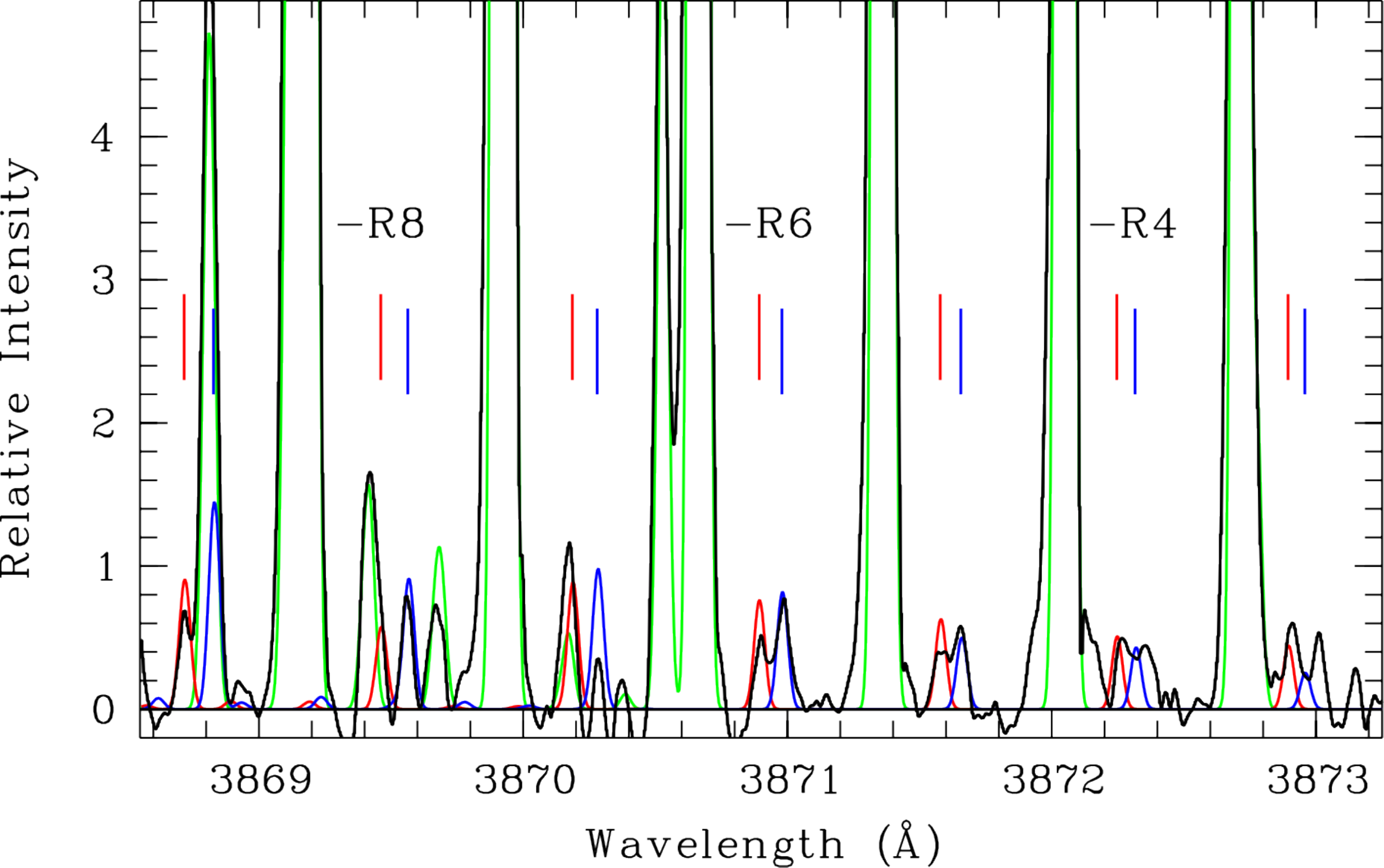}
\caption{UVES spectra (black line) compared to synthetic spectra of isotopic species: $^{12}$C$^{14}$N (green line), $^{12}$C$^{15}$N (red line), and $^{13}$C$^{14}$N (blue line). The synthetic spectra are computed with the adopted isotopic abundances. The lines of $^{12}$C$^{15}$N and $^{13}$C$^{14}$N are identified by the red ticks and the blue ticks, respectively, and a few R lines are indicated by their quantum number. 
}
\label{cn}
\end{figure} 

\section{Results}

\begin{figure}[!hb]
\includegraphics[ width=9cm]{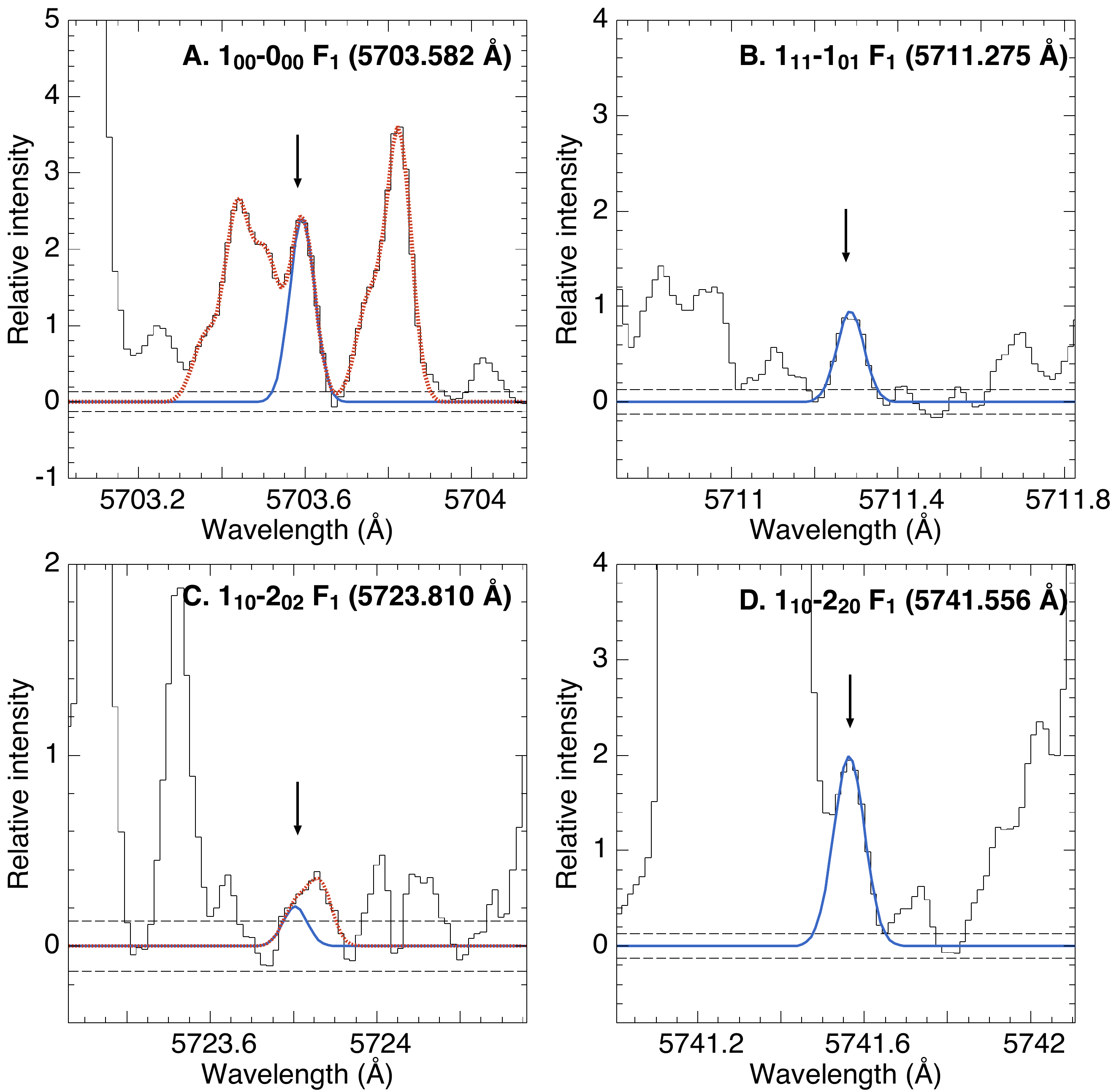}
\caption{Four measured $^{15}$NH$_2$ lines of comet C/2015 ER61. Blue solid lines and black arrows show the $^{15}$NH$_2$ lines. Red dotted lines in panels A and C indicate the results of ad hoc fitting with multiple Gaussian functions. Gray dashed lines indicate the $\pm$1$\sigma$ error of the continuum levels. 
}
\label{nh2}
\end{figure} 

In Figure~\ref{cn}, we present the combined UVES spectrum of comet ER61, taken at $r=$1.13 and $r=$1.11 au, shortly after the outbursting event. Only the R lines of the B-X (0, 0) band (i.e., shortward of 3875 \AA) are used since the P lines of the three isotopologs of CN are strongly blended. The $^{12}$C/$^{13}$C and $^{14}$N/$^{15}$N ratios are obtained simultaneously by fitting the CN B-X (0, 0) band with theoretical models. More details of our models are described in \cite{Manfroid:2009}. Based on measurements of the CN band, we derived $^{12}$C/$^{13}$C=100 $\pm$ 15 and $^{14}$N/$^{15}$N=130 $\pm$ 15. The uncertainties of our measurements are not dominated by random errors, but come from the subtraction of the underlying continuum (i.e.,\ the dust continuum plus the faint wings of the strong lines) and inaccuracies in the models we used. We estimated that they roughly correspond to 2$\sigma$ errors, beyond which reasonable fits of the isotopic species are not acceptable. We also attempted to measure $^{16}$O/$^{18}$O using the OH ultraviolet bands at 3063 \AA\ (0, 0) and 3121 \AA\ (1, 1). Some $^{18}$OH lines were marginally detected, but the signal-to-noise ratio of these lines is too low to derive a reliable $^{16}$O/$^{18}$O estimate.

In addition, we also derived the nitrogen isotopic ratio using NH$_2$ lines around 5700 \AA. The advantage of using NH$_2$ over CN is that NH$_2$ is the dominant photodissociation product of NH$_3$, which is directly incorporated into the nucleus, whereas there are multiple possible parent molecules of CN besides HCN. The methodology and assumptions for measuring the $^{14}$N/$^{15}$N ratio of NH$_2$ are explained in \cite{Shinnaka:2016}. We derive the $^{14}$N/$^{15}$N ratio of NH$_2$ of 140 $\pm$ 28, which is consistent with the nitrogen isotopic ratio of CN within the uncertainties. The uncertainty of the NH$_2$ measurements consists of both random and systematic errors, where the random errors come from flux measurements of the isotopolog lines and the continuum subtraction and systematic errors come from theoretical assumptions. The two key assumptions are, first, the same transitions in $^{14}$NH$_2$ and $^{15}$NH$_2$ have consistent transition probabilities and, second, NH$_3$ and $^{15}$NH$_3$ have similar photodissociation rates to produce the two isotopologs. 

\section{Discussion} 
ER61 developed a dust coma when it was near 8 au from the Sun, where the surface temperature is too cold for water ice to sublimate. This suggests that the coma was driven by more volatile species such as CO or CO$_2$. However, if these super volatiles were near the surface, the comet would have been active beyond 10 au, which is not the case for ER61. There are two possibilities to account for the distant activity: either these volatiles were buried beneath a refractory layer or they were trapped in amorphous water ice, which is thermodynamically unstable and converts exothermically to the crystalline form \citep{Prialnik:2004}. However, no near-infrared spectroscopy of ER61 is available and the physical state of water ice in ER61 is unknown. Although we cannot rule out the possibility that amorphous-to-crystalline transition of water ice is responsible for the distant activity of ER61, the analysis of the activity pattern of ER61 and thermal models indicates that CO or CO$_2$ on this comet were depleted down to depths of 6.9\,m and 0.4\,m, respectively \citep{Meech:2017}. Assuming water release is controlled only by insolation, \citet{Meech:2017} have predicted that the comet would show relatively moderate activity when approaching the sun. This prediction is consistent with the constant gas production rates measured from 2.3 au to 1.3 au from the Sun by TRAPPIST-South telescope just before the outburst in 2017 March \citep{Jehin:2017}. The orbital characteristics of ER61 suggest that it is not a dynamically new comet. Given that it had entered the inner solar system before, the activity pattern of ER61 indicates that this comet has developed an aging surface that consists of an insulating porous dust layer. It has been previously shown that such a dust layer is efficient in protecting the subsurface materials from heat, solar wind, or minor impacts \citep{Schorghofer:2008,Guilbert:2015}. The outburst in 2017 April, thus, is of particular importance because it enabled us to sample the well-preserved subsurface materials released into the coma via the outburst, which are otherwise hard to access. 

\begin{figure}[b]
\includegraphics[width=9.0cm]{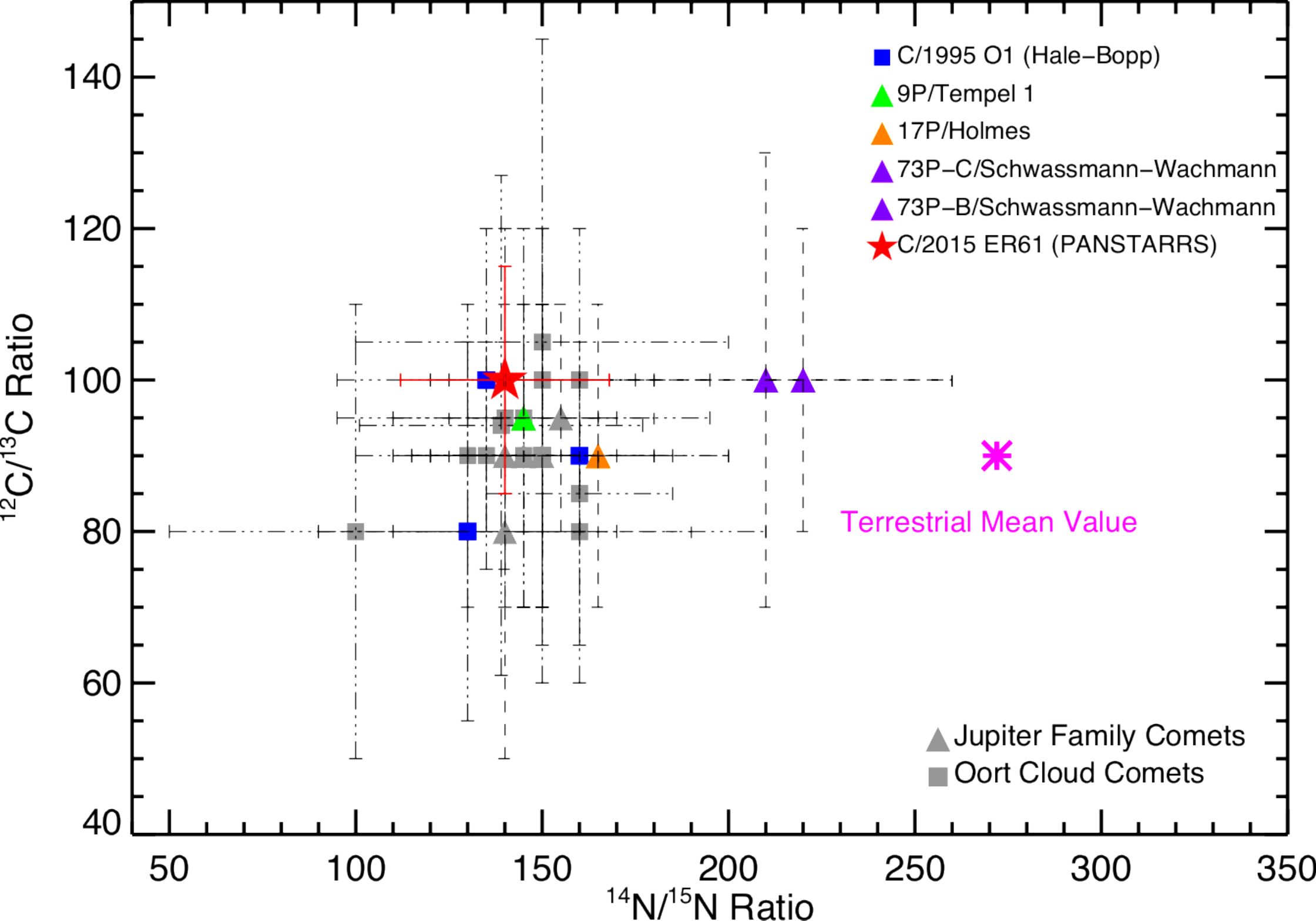}
\caption{Comparison between the isotopic ratios of ER61 with those measured for other comets shows that ER61 is typical of other comets. Data are from \cite{Manfroid:2009}, \cite{Bockelee-Morvan:2015}, and references therein. The measurements of OCCs are shown as squares and JFCs as triangles. Isotopic ratios for C/1995 O1 (Hale-Bopp) are measured three times at different heliocentric distances. The terrestrial isotopic ratios are from \cite{Anders:1989}. 
}
\label{C$_2$}
\end{figure} 

Shown as the red star in Figure 4, our measurements of $^{14}$N/$^{15}$N and $^{12}$C/$^{13}$C in ER61 do not exhibit significant difference from those of other comets, even though the measured materials are expected to be pristine. Similarly, \cite{Bockelee-Morvan:2008} observed comet 17P/Holmes soon after its remarkable outburst in 2007 and measured $^{14}$N/$^{15}$N = 139 $\pm$ 26 in HCN, while \cite{Manfroid:2009} measured $^{14}$N/$^{15}$N = 165 $\pm$ 40 in CN, which are also unremarkable compared to other comets. Also, there is no evidence that the subsurface material released from comet 9P/Tempel 1, as a result of the Deep Impact event, was isotopically different from the surface material \citep{Jehin:2006}. The only outliers that have possibly different isotopic ratios are the fragments of the split nucleus of 73P/Schwassmann-Wachmann 3 (hereafter P/SW3). \cite{Shinnaka:2011} noted that the measured ortho-to-para abundance ratios (OPR) of 1.0 in NH$_3$ correspond to a high nuclear spin temperature. Other observations show that P/SW3 is strongly depleted in C$_2$ and NH$_2$ relative to CN and OH \citep{Schleicher:2011}. The high $^{14}$N/$^{15}$N ratio in CN and the extreme depletions consistently suggest that P/SW3 formed under relatively warm conditions. This is an interesting finding because P/SW3 is a Jupiter family comet (JFC) and should have formed at 30 - 50 au according to the classical picture \citep{Levison:1997}.

Before the significance of the migration of the giant planets was recognized, JFCs and Oort cloud comets (OCCs) were thought to have formed in distinct regions, where JFCs formed beyond Neptune and OCCs formed between giant planets \citep{Dones:2004}. Except for P/SW3, we see little if any systematic difference between JFCs and OCCs, even when freshly exposed interior materials were measured. In terms of chemical composition, surveys of cometary volatiles have also shown no systematic difference between JFCs and OCCs, but that significant variations are present within each group \citep{mumma:2011,AHearn:2012}. 

As growing observational evidence reveals that JFCs and OCCs formed within the same broad region, dynamical simulations, in the framework of planetary migration, also suggest that the Oort cloud and scattered disk are derived from a common parent population, i.e., the primordial trans-Neptunian disk \citep{Brasser:2013}. Recent measurements of HDO/H$_2$O in 67P \citep{Altwegg:2015} by the ROSINA mass spectrometer on board the Rosetta spacecraft, also suggest that JFCs and OCCs formed in largely overlapping regions where the giant planets are today \citep{AHearn:2017}. However, the proportions of comets formed at different heliocentric distances are not yet clear \citep{Dones:2015}.

\cite{Furi:2015} studied the distribution of N isotopes in the solar system and found that it follows a rough trend with increased $^{15}$N abundances at larger radial distances from the Sun. Nitriles detected in cometary comae so far are consistently enriched in $^{15}$N by a factor of $\sim$3 relative to the protosolar nebula and a factor of 1.8 relative to the terrestrial planets \citep{Furi:2015}. The significant enrichment in $^{15}$N is unlikely to be due to primordial nucleosynthetic heterogeneities, but rather a result of isotope fractionation processes during the formation of the solar system \citep{Furi:2015}. Possible processes include ion-molecule reactions in the interstellar medium under cold temperature with sufficient density \citep{charnley:2002} or photodissociation of N$_2$ by UV light from the proto-Sun or nearby stars \citep{muskatel:2011}. 

Unlike nitrogen, the available carbon isotope ratios of comets, including the newly derived $^{12}$C/$^{13}$C ratio of ER61, are consistent with the solar system abundance ratio of 90 $\pm$ 10 \citep{Wyckoff:2000}. The in situ measurements of the CO$_2$ isotopologs in the coma of 67P/Churyumov-Gerasimenko, taken by the Double Focusing Mass Spectrometer (DFMS) of the ROSINA, find that $^{12}$C/$^{13}$C = 84 $\pm$ 4 \citep{Hassig:2017}. The DFMS measurements show a slight enrichment in $^{13}$C in terms of the $^{12}$C/$^{13}$C ratio compared to an average $^{12}$C/$^{13}$C ratio of 91.0 $\pm$ 3.6 \citep{Manfroid:2009,Bockelee-Morvan:2015}. However, given the large uncertainties of other ground-based observations, this difference seen in the DFMS measurements of 67P is not statistically significant. Our observations and previous studies consistently suggest that comets have formed in the Sun's protoplanetary disk and inherited the $^{12}$C/$^{13}$C ratio from the original protosolar nebula \citep{Hassig:2017} and the chemical fractionation did not significantly change the ratios after the molecules formed in the protosolar cloud \citep{Wyckoff:2000}.

\section{Conclusions}
We performed high-resolution spectroscopy on the outbursting comet C/2015 ER61 on 2017 April 13 and 17, approximately one week after the outburst event. Our main results are summarized as follows:

First, we derived $^{14}$N/$^{15}$N and $^{12}$C/$^{13}$C in CN to be 130 $\pm$ 15 and 100 $\pm$ 15, respectively. In addition, we derived $^{14}$N/$^{15}$N = 140 $\pm$ 28 in NH$_2$.
Second, although it is likely that the fresh subsurface materials of ER61 were analyzed, both the nitrogen and carbon isotopic ratios of this comet do not deviate significantly from those of other comets. 
Finally, this work and previous studies consistently suggest that JFCs and OCCs originated in largely overlapping regions beyond proto-Neptune.

\begin{acknowledgements}  
Based on observations collected at the European Organisation for Astronomical Research in the Southern Hemisphere under ESO programme 299.C-5006(A). TRAPPIST-South is funded by the Belgian Fund for Scientific Research (Fond National de la Recherche Scientifique, FNRS) under the grant FRFC 2.5.594.09.F. D.H. and E.J. are Belgian FNRS Senior research Associates, and M.G. is FNRS research Associate. YS was supported by Grant-in-Aid for JSPS Fellows Grant No. 15J10864. KJM and JVK acknowledge support from NSF AST-1617015. The authors would like to thank the referee, Anita Cochran, for her thoughtful and constructive comments. 
\end{acknowledgements}



\end{document}